\documentclass[a4paper,dvips,12pt]{article}
\input{axodraw.sty}
\usepackage{amsmath,amssymb,exscale}   
\usepackage{array,multicol}
\usepackage{afterpage,float,flafter}   
\usepackage{epsfig,rotating,pifont,fancybox}   
\usepackage{cite}
\makeatletter   
\@addtoreset{equation}{section}   
\makeatother   
   
\newcommand{\bea}{\begin{eqnarray}}   
\newcommand{\eea}{\end{eqnarray}}   
   
\newcommand{\NPB}[3]{\emph{ Nucl.~Phys.} \textbf{B#1} (#2) #3}   
\newcommand{\PLB}[3]{\emph{ Phys.~Lett.} \textbf{B#1} (#2) #3}   
\newcommand{\PRD}[3]{\emph{ Phys.~Rev.} \textbf{D#1} (#2) #3}   
\newcommand{\PRL}[3]{\emph{ Phys.~Rev.~Lett.} \textbf{#1} (#2) #3}

\newcommand{\JHEP}[3]{\emph{ JHEP} \textbf{#1} (#2) #3}

\title{   
\vspace*{-0.8cm}   
\begin{flushright}   
\normalsize{      
IEM-FT-228/02\\
\texttt{hep-th/0210181}}\\ 
\end{flushright}    
\vspace{1cm}
\Large{\sc Brane-assisted Scherk-Schwarz supersymmetry breaking in
orbifolds~\footnote{Work supported in part by CICYT, Spain, under
contracts FPA 2001-1806 and FPA 2002-00748, by EU under contracts
HPRN-CT-2000-00152 and HPRN-CT-2000-00148, and by NSF under grants
P420D3620414350 and P4203620434350.}}
\vspace*{.5cm} \author{\large {\sc A.~Delgado~$^a$,
G.~v.~Gersdorff~$^b$ and M.~Quir{\'o}s~$^b$}\\ \\
\emph{~$^a$Department of Physics and Astronomy, Johns Hopkins
University}\\ \emph{3400 North Charles Street, Baltimore, MD
21218-2686, USA}\\ \emph{~$^b$Instituto de Estructura de la Materia (CSIC),
Serrano 123}\\ \emph{E-28006-Madrid, Spain}} } \date{}
\begin{document}
\maketitle
\thispagestyle{empty}
\vspace*{.5cm}

\begin{abstract}\noindent
We have analyzed the interplay between Scherk-Schwarz supersymmetry
breaking and general fermion (gaugino or gravitino) mass terms
localized on the fixed point branes of the $S^1/\mathbb{Z}_2$
orbifold.  Analytic solutions for eigenfunctions and eigenvalues are
found in all cases.  All results are checked by numerical calculations
that make use of regularized $\delta$-functions.  Odd and generically
also even fermions are discontinuous at the brane fixed points, but in
all cases the combination that couples to the brane is continuous.
For CP-even brane mass terms supersymmetry restoration can take place
when their effects are cancelled by those of Scherk-Schwarz
compactification. However such a cancellation can not occur for CP-odd
brane mass terms.
\end{abstract}
\vspace{2.cm}   
   
\begin{flushleft}   
October 2002 \\   
\end{flushleft}
%%%%%%%%%%%%%%%%%%%%%%%%%%%%%%%%%%%%%%%%%%%%%%%%%%%%%%   
\newpage
\section{\sc Introduction}
\label{introduction}

Extra dimensions at the TeV scale~\cite{ignas} have become a popular
feature in nowadays model building. They can provide possible
explanations to some of the present problems within the minimal
supersymmetric extension of the Standard Model (MSSM) like
supersymmetry breaking, the flavour structure, electroweak symmetry
breaking, or even unification with gravity. In particular, one of the
most appealing possibilities to break supersymmetry in theories with
extra dimensions is the Scherk-Schwarz (SS) mechanism~\cite{SS}, based
on non-trivial twists that generate a mass term through
compactification~\cite{matter,UV}. This mechanism appears to be
natural in any locally supersymmetric five-dimensional (5D) theory,
since it has been proven that the SS mechanism can be seen as a
spontaneous breaking of the $SU(2)_R$ via some off-shell auxiliary
fields~\cite{SSH1,SSH2,radionF}. Being spontaneous this kind of supersymmetry
breaking is quite desirable since it will not generate any quadratic
divergences~\cite{one}.

The most common way of compactifying extra-dimensional theories is by using
orbifolds~\cite{orbifolds} which allow for chirality and partial
breaking of supersymmetry; this is needed because theories in extra
dimensions are in general non-chiral and have extended
supersymmetry. Orbifolds introduce fixed points in the theory, where
translational invariance is broken.  
Therefore terms localized at those points can arise without spoiling
the higher dimensional Poincare invariance in the bulk.  This type of
localized interactions can be used to break supersymmetry dynamically,
e.g.~via gaugino condensation, or by any other kind of mechanism that
could generate a brane mass term for gravitinos. The possibility of
supersymmetry restoration between the SS mechanism and brane localized
supersymmetry breakdown was already proposed in the context of the
heterotic Horava-Witten M-theory~\cite{AQ}.  Also, this kind of
interactions have been studied in the past in the context of
generalized SS mechanism~\cite{fabio1}, since these terms completely
change the spectrum and moreover any fermion mass term localized on a
brane will generate discontinuities in the wavefunctions. The aim of
this paper is to generalize previous
results~\cite{fabio1,fabio2,SSH2,nilles,feruglio} and make a
systematic analysis of a very general class of models that could shed
light on the mechanism of supersymmetry breaking.

In this paper we will study a 5D theory compactified on $S^1/
\mathbb{Z}_2$ with SS supersymmetry breaking plus the most general
fermion mass terms on the brane, which do not necessarily have to
respect $SU(2)_R$ since the latter is broken by the
compactification. We will perform a fully analytic study of both the
CP-even and the CP-odd brane mass terms and find out the cases where
there are massless states in the spectrum, corresponding to situations
where there is cancellation of supersymmetry breaking between the
different sources. We will also show how the discontinuities of the
fields are treated consistently and are independent of any
regularization of the brane. We will perform all explicit calculations
for the case of gauginos that are symplectic Majorana spinors
transforming as doublets with respect to $SU(2)_R$. Our results will
be straightforwardly generalized to the case of symplectic Majorana 5D
gravitinos that also transform as doublets of $SU(2)_R$. All our
analytic results have been numerically checked using regularized
$\delta$-functions.

The layout of the paper is as follows. In section 2 we will present
the model with both the bulk Lagrangian plus possible mass terms
localized on the brane that will be classified according to its
transformation properties under the four-dimensional CP-invariance.
Section 3 will be devoted to the full analysis of the CP-even brane
whereas section 4 contains the CP-odd case. Finally our conclusions
will be drawn on section 5.

\section{\sc Fermion bulk Lagrangian and brane-mass terms}
We will consider a five-dimensional model compactified on the orbifold
$S^1/\mathbb{Z}_2$ where $\mathbb{Z}_2$ transforms the fifth
coordinate $x^5\equiv y$ into $-y$. Fermions in five dimensions can be
described by a pair of symplectic Majorana spinors $\lambda^i$ that
transform as a doublet under $SU(2)_R$ and such that

\begin{equation}
\label{symplectic}
\lambda^i=\left(
\begin{array}{c}
\lambda^i_L\\
\epsilon^{ij}\bar{\lambda}_{L\,j}
\end{array}
\right)
\end{equation}
where $\lambda^i_L$ are the two-component spinors $\lambda_L\equiv
\lambda_\alpha$ and $\bar{\lambda}_L\equiv
\bar\lambda^{\stackrel{\cdot}{\alpha}}$. Under $\mathbb{Z}_2$ the
transformation law of $\lambda$ is
\begin{equation}
\label{law}
\lambda(y)\to \mathcal{P}_\lambda \sigma^3 i\gamma^5 \lambda(-y)
\end{equation}
where $\mathcal{P}_\lambda=\pm$ is the intrinsic parity of the spinor
$\lambda$, the matrix $\sigma^3$ is acting on $SU(2)_R$ indices,
$\gamma^5=diag(-i,i)$ and we are using the metric
$\eta_{MN}=diag(1,-1,-1,-1,-1)$. Then for a spinor with intrinsic
parity $\mathcal{P}_\lambda=+$ it turns out that $\lambda^1_L(y)$ is
an even and $\lambda^2_L(y)$ an odd two-component spinor.

We will find it convenient to define 4D Majorana spinors as
\begin{equation}
\label{Majorana}
\psi^i=\left(
\begin{array}{c}
\lambda^i_L\\
\bar{\lambda}_{L\,i}
\end{array}
\right)
\end{equation}
although obviously they do not transform irreducibly under $SU(2)_R$
nor are they covariant under the 5D Lorentz
group. Under $\mathbb{Z}_2$ parity $\psi^1(y)$ is even and $\psi^2(y)$
odd.

The Lagrangian for $\lambda$ in the bulk contains
\begin{equation}
\label{bulk}
\mathcal{L}_{\rm bulk}=i\, \bar\lambda \gamma^M \mathcal{D}_M \lambda
\end{equation}
where $\mathcal{D}_M$ is the covariant derivative with respect to
local Lorentz and gauge transformations and summation over the
$SU(2)_R$ indices is always understood. The kinetic Lagrangian is
obtained from the 4D covariant derivative $\mathcal{D}_\mu$ as
\begin{equation}
\label{kin}
\mathcal{L}_{\rm kin}=i\, \bar\lambda \gamma^\mu \partial_\mu
\lambda=i\, \bar\psi \gamma^\mu \partial_\mu \psi
\end{equation}
while the KK-mass bulk term comes from the $\partial_5$ term in
$\mathcal{D}_5$ as
\begin{equation}
\label{KKmass}
\mathcal{L}_{\rm KK-mass}=i\, \bar\lambda \gamma^5
\partial_5\lambda=\bar\psi_2\partial_5
\psi^1-\bar\psi_1\partial_5\psi^2
\end{equation}

In the off-shell formulation of 5D supergravity~\cite{zucker,kugo} it was
proven that $\mathcal{D}_M$ also contains the covariant derivative
with respect to local $SU(2)_R$ tansformations,
$-i\vec{\sigma}\vec{V}_M$, where $\vec{V}_M$ are auxiliary fields in
the minimal supergravity multiplet. This fact was used in
Refs.~\cite{SSH1,SSH2} to interpret the Scherk-Schwarz SS breaking
of supersymmetry as the background configuration
$\langle\vec{\sigma}\vec{V}_M\rangle=\delta_{M5}\sigma^2 \langle
V_5^2\rangle\equiv \delta_{M5}\sigma^2 \omega$~\footnote{We are using
units where the compactification radius $R\equiv 1$. Restoring the
$R$-dependence at any stage should be a trivial task.}. This gives
rise to the SS-Lagrangian
\begin{equation}
\label{SS}
\mathcal{L}_{\rm SS}=i\, \omega \bar\lambda\gamma^5\sigma^2
\lambda=-\omega\left(\bar\psi_1\psi^1+\bar\psi_2\psi^2\right)
\end{equation}
The bulk Lagrangian can then be written as
\begin{equation}
\label{bulk2}
\mathcal{L}_{\rm bulk}=i\,\bar\psi_i\gamma^\mu\partial_\mu
\psi^i+
\bar\psi_2\partial_5\psi^1-\bar\psi_1\partial_5\psi^2-
\omega(\bar\psi_1\psi^1+\bar\psi_2\psi^2)
\end{equation}

In the brane the different possible $\mathbb{Z}_2$-even mass terms can
be either $CP$-even or $CP$-odd. $CP$-even mass terms on the brane are
provided by the following contributions
\begin{equation}
\label{CP1}
\bar\lambda\,\sigma^2\gamma^5\lambda=-(\bar\psi_1\psi^1+\bar\psi_2\psi^2)
\end{equation}
and
\begin{equation}
\label{CP2}
\bar\lambda\,\sigma^1\lambda=-(\bar\psi_1\psi^1-\bar\psi_2\psi^2)
\end{equation}
or an arbitrary linear combination of (\ref{CP1}) and (\ref{CP2}). We
will then consider possible mass terms localized at one of the branes,
e.g. $y=0$, as
\begin{equation}
\label{CPeven}
\mathcal{L}_{\rm brane,\,
CP-even}=-2\Lambda\delta(y)\left[\bar\psi_1\psi^1+\rho\,\bar\psi_2\psi^2\right]
\end{equation}
where $\rho$ is an arbitrary real parameter.

$CP$-odd mass terms on the brane can be provided by the contributions
\begin{equation}
\label{noCP1}
\bar\lambda\,\sigma^1\gamma^5\lambda=-
(\bar\psi_1\gamma^5\psi^1-\bar\psi_2\gamma^5\psi^2)
\end{equation}
and
\begin{equation}
\label{noCP2}
\bar\lambda\,\sigma^2\lambda=\bar\psi_1\gamma^5\psi^1+\bar\psi_2\gamma^5\psi^2
\end{equation}
or any arbitrary linear combination of (\ref{noCP1}) and (\ref{noCP2}). We
can then consider the brane Lagrangian
\begin{equation}
\label{CPodd}
\mathcal{L}_{\rm brane,\,
CP-odd}=-2\Lambda\delta(y)\left[\bar\psi_1\gamma^5\psi^1-
\rho\,\bar\psi_2\gamma^5\psi^2\right]
\end{equation}
where $\rho$ is an arbitrary real parameter.

Five-dimensional theories where brane mass terms (\ref{CP1}) and
(\ref{CP2}), or (\ref{noCP1}) and (\ref{noCP2}), appear can be found
in the recent literature. On the other hand solving theories where the
brane Lagrangians (\ref{CPeven}) and (\ref{CPodd}) interfere with the
Scherk-Schwarz breaking has been done only for some cases. In
particular the Lagrangian (\ref{CPeven}) for
$\rho=0,1$~\cite{fabio1,nilles} and the Lagrangian (\ref{CPodd}) for
$\rho=0$~\cite{SSH2}. In the following sections we will analyze the
general cases both for (\ref{CPeven}) and (\ref{CPodd}).

\section{\sc CP-even brane mass terms}
The Lagrangian in this case is given by (\ref{bulk2}) and
(\ref{CPeven}), i.e.
\begin{equation}
\label{Lageven}
\mathcal{L}_{\rm CP-even}=i\,\bar\psi_i\gamma^\mu\partial_\mu \psi^i+
\bar\psi_2\partial_5\psi^1-\bar\psi_1\partial_5\psi^2-
\omega(\bar\psi_1\psi^1+\bar\psi_2\psi^2)-
2\Lambda\delta(y)\left[\bar\psi_1\psi^1+\rho\,\bar\psi_2\psi^2\right]
\end{equation}

The equations of motion (EOM) from the Lagrangian (\ref{Lageven}) are
given by
\begin{align}
\label{EOM0}
\left(-i\, \gamma^\mu\partial_\mu+\omega\right)\psi^1_n+\partial_y
\psi^2_n+2\phantom{\rho}\Lambda\delta(y)\psi^1_n=&0\nonumber\\
\left(-i\, \gamma^\mu\partial_\mu+\omega\right)\psi^2_n-\partial_y
\psi^1_n+2\rho\,\Lambda\delta(y)\psi^2_n=&0
\end{align}
We can now decompose the fields $\psi^i_n(x,y)$ in plane waves as
$\psi^i_n(x,y)=\exp(-ip\cdot x)\psi^i_n(p,y)$ where $p^2=M_n^2$, $M_n$
being the mass eigenvalues and $\psi^i_n(p,y)$ the mass
eigenfunctions. In the rest frame $p^\mu=(M_n,\vec{0})$
Eq.~(\ref{EOM0}) is written as
\begin{align}
\label{EOM1}
\left(-M_n\gamma^0+\omega\right)\psi^1_n+\partial_y
\psi^2_n+2\phantom{\rho}\Lambda\delta(y)\psi^1_n=&0\nonumber\\
\left(-M_n\gamma^0+\omega\right)\psi^2_n-\partial_y
\psi^1_n+2\rho\,\Lambda\delta(y)\psi^2_n=&0\ .
\end{align}
where $\psi^i_n=\psi^i_n(M_n,y)$ are the mass eigenstates in the rest
frame.

We define now Majorana spinors as
\begin{equation}
\label{Majorana2}
\psi^i_n=\left(
\begin{array}{c}
\phi^i_n(y)\ \xi_n  \vspace{.25cm}\\ 
\bar\phi_{i\,n}(y)\ \xi_n
\end{array}
\right)
\end{equation}
where $\xi_n$ are constant real spinors and $\phi^i_n(y)$ complex functions.
We can then write the EOM (\ref{EOM1}) as
\begin{align}
\label{EOM2}
-M_n\bar\phi^1_n+\omega\phi^1_n+\partial_y
\phi^2_n+2\phantom{\rho}\Lambda\delta(y)\phi^1_n=&0\nonumber\\
-M_n\bar\phi^2_n+\omega\phi^2_n-\partial_y
\phi^1_n+2\rho\,\Lambda\delta(y)\phi^2_n=&0
\end{align}
where the constant two-component spinors $\xi_n$ factorize out of the
EOM (\ref{EOM2}) and play no role in what follows.

An explicit analytic solution of equations (\ref{EOM2}) with real
functions $\bar{\phi}^i_n(y)=\phi^i_n(y)$ can be obtained in the
region $-2\pi<y<2\pi$, for initial conditions 
\begin{align}
\label{initialeven}
\phi^1_n(0)=&1\nonumber\\
\phi^2_n(0)=&0,
\end{align}
as
\begin{equation}
\label{solutioneven}
\left(
\begin{array}{c}
\phi^1_n(y)\vspace{.25cm}\\ 
\phi^2_n(y)
\end{array}
\right)=\sqrt{\frac{1+t_\rho^2\left(\Lambda\varepsilon(y)\right)}
{1+\rho\ t_\rho^2\left(\Lambda\varepsilon(y)\right)}}
\left(
\begin{array}{c}
\cos\left[(M_n-\omega)y-
\arctan\,t_\rho\left(\Lambda\varepsilon(y)\right)\right] \vspace{.25cm}\\
\sin\left[(M_n-\omega)y-
\arctan\,t_\rho\left(\Lambda\varepsilon(y)\right)\right]
\end{array}
\right)
\end{equation}
where $\varepsilon(y)={\rm sign}(y)$ is the sign function and we use
the notation,
\begin{equation}
\label{notation}
t_\rho(x)= \tan(\sqrt{\rho}\,x)/\sqrt{\rho}
\end{equation}
For the typical values of $\rho=1,0,-1$ this function is given by:
\begin{align}
\label{values1}
 t_1(x)=&\tan(x)\\
\label{values0}
 t_0(x)=&x\\
\label{values-1}
 t_{-1}(x)=&\tanh(x)
\end{align}

In the region $-2\pi<y<2\pi$ the functions (\ref{solutioneven}) are
periodic provided the condition
\begin{equation}
\label{periodiceven}
M_n=n+\omega+\frac{1}{\pi}\arctan\, t_\rho(\Lambda)
\end{equation}
holds. Notice that supersymmetry can be restored if $\omega$ and
$\Lambda$ are such that $\omega\pi+\arctan\, t_\rho(\Lambda)=n\pi$,
where $n$ is any integer number in which case there is always a
massless mode in the spectrum.

The solution (\ref{solutioneven}) near the brane, i.e. for
$|y|<\epsilon$ where $\epsilon\to 0$, is given by
\begin{equation}
\label{evenbrane}
\left(
\begin{array}{c}
\phi^1_n(y) \vspace{.25cm}\\
\phi^2_n(y)
\end{array}
\right)=\frac{1}{\sqrt{
1+\rho\ t_\rho^2\left(\Lambda\varepsilon(y)\right)}}
\left(
\begin{array}{c}
1 \vspace{.25cm} \\
-t_\rho\left(\Lambda\varepsilon(y)\right)
\end{array}
\right)
\end{equation}
From (\ref{evenbrane}) we see that $\phi^2_n$ is an odd function that
makes a jump at the brane $\Delta\phi^2_n$ given by
\begin{equation}
\label{jumpeven}
\left|\Delta\phi^2_n\right|=\frac{2 t_\rho(\Lambda)}{\sqrt{
1+\rho\ t_\rho^2\left(\Lambda\right)}}
\end{equation}
while $\phi^1_n$ has a discontinuity on the brane
\begin{equation}
\label{disceven}
\lim_{y\to 0}\phi^1_n(y)=\frac{1}{\sqrt{
1+\rho\ t_\rho^2(\Lambda)}}
\end{equation}
Which in general is different from $\phi^1_n(0)=1$. Only for the
particular case $\rho=0$ is the function $\phi^1_n(y)$ continuous. However one can check that near the brane
\begin{equation}
\label{conteven}
\lim_{y\to
0}\left[(\phi^1_n(y))^2+\rho\,(\phi^2_n(y))^2\right]=
(\phi^1_n(0))^2+\rho\,(\phi^2_n(0))^2
\end{equation}
and so the term in the Lagrangian (\ref{Lageven}) coupled to the brane
is continuous.

\begin{figure}[htb]
%\vspace{.8cm}
\centering
\epsfig{file=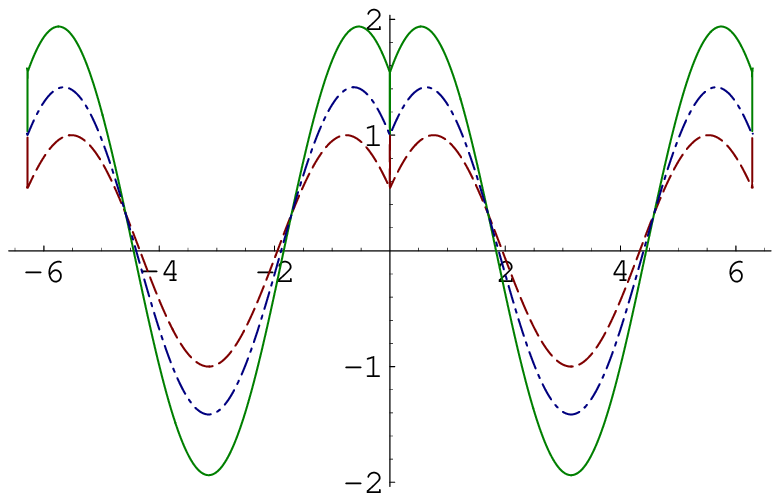,width=.45\linewidth}
\epsfig{file=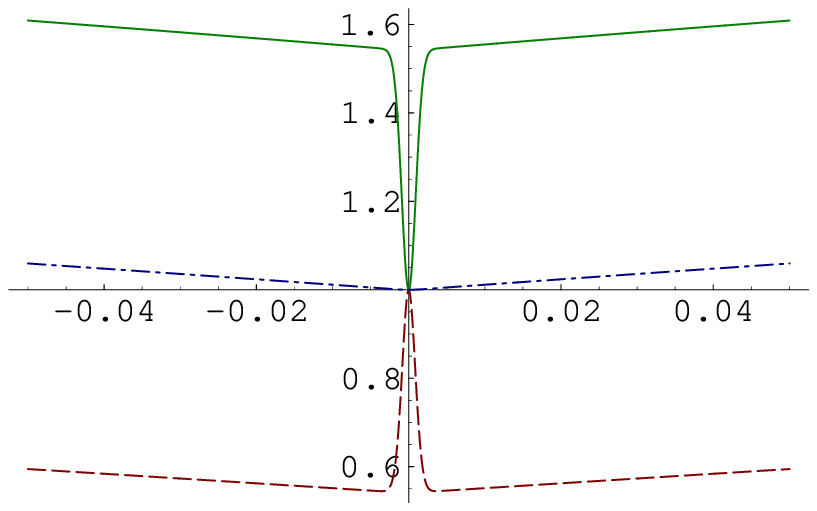,width=.45\linewidth}
\epsfig{file=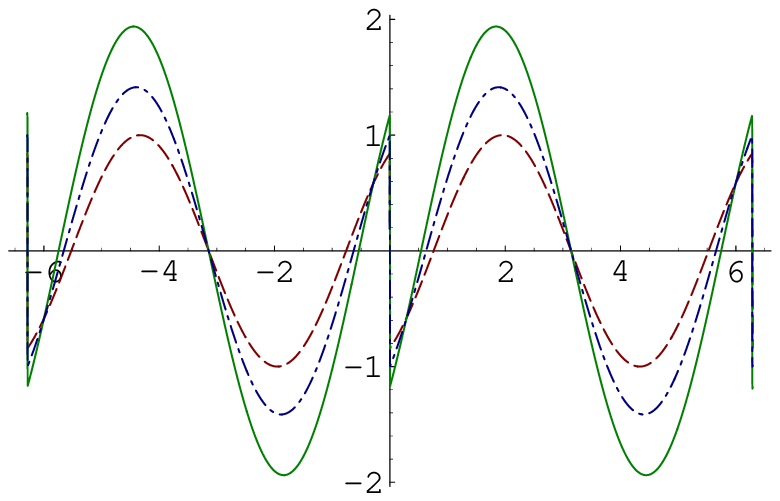,width=.45\linewidth}
\epsfig{file=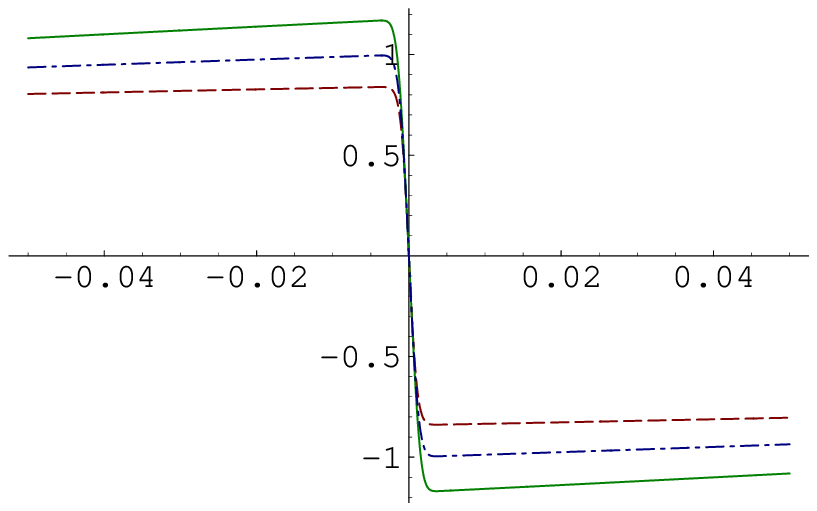,width=.45\linewidth}
\caption{\it 
The numerical solution of Eq.~(\ref{EOM2}) for $\Lambda=1$ and
$\omega=0.2$. We show three different values of $\rho$,
$\rho=-1$ (solid line), $\rho=1$ (dashed line) and $\rho=0$
(dashed-dotted line). In the right plot we display in detail the
different discontinuities.
}
\label{cpevenplot}
\end{figure}

\begin{figure}[htb]
%\vspace{.8cm}
\centering
\epsfig{file=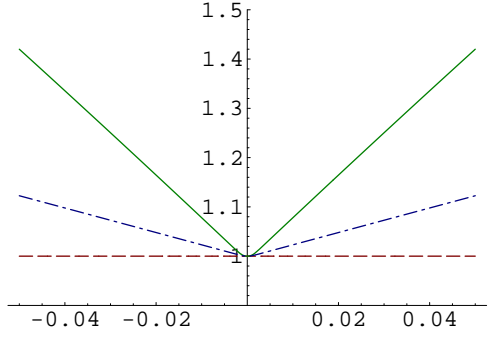,width=.45\linewidth}
\caption{\it 
The quantity $(\phi^1_n(y))^2+\rho\,(\phi^2_n(y))^2$ taken from the
numerical solution with a gaussian-regularized $\delta$-function near
the origin. Again, we diplay the cases $\rho=-1$ (solid line),
$\rho=1$ (dashed line) and $\rho=0$ (dashed-dotted line). In all cases
the quantity is continous in accordance with the analytical solution.
}
\label{cpevencont}
\end{figure}

In Fig.~\ref{cpevenplot} we present a numerical solution to
Eq.~(\ref{EOM2}) obtained by regularizing the $\delta$-function with a
gaussian of width $\sigma=10^{-3}$ (we have checked our results using
other regularizations as well).  The solutions precisely coincide with
the analytic ones, Eqs.~(\ref{solutioneven}) and
(\ref{periodiceven}). We display in detail the discontinuities found
in Eq.~(\ref{jumpeven}) and Eq.~(\ref{disceven}). Furthermore we plot
in Fig.~\ref{cpevencont} the combination
$(\phi^1_n(y))^2+\rho\,(\phi^2_n(y))^2$.  Note that the scales used in
Fig.~\ref{cpevencont} and in Fig.~\ref{cpevenplot} (right panel) are
identical and the continuity of
$(\phi^1_n(y))^2+\rho\,(\phi^2_n(y))^2$ is explicit.

If we replace $\delta(y)\to \sum_k \delta(y-2\pi k)$ in
(\ref{Lageven}) the solution (\ref{solutioneven}) can be generalized
to one periodic for $y\in\mathbb{R}$ as
\begin{equation}
\label{pereven}
\left(
\begin{array}{c}
\phi^1_n(y) \vspace{.25cm}\\
\phi^2_n(y)
\end{array}
\right)=\sqrt{\frac{1+t_\rho^2\left(\Lambda\beta(y)\right)}
{1+\rho\ t_\rho^2\left(\Lambda\beta(y)\right)}}
\left(
\begin{array}{c}
\cos\left[(M_n-\omega)y-\widetilde{\eta}(y)\right]  \vspace{.25cm}\\
\sin\left[(M_n-\omega)y-\widetilde{\eta}(y)
\right]
\end{array}
\right)
\end{equation}
where
\begin{equation}
\label{beta}
\beta(y)=
\lim_{\kappa\rightarrow\infty}
\sum_{k=-\kappa}^\kappa
(-)^k\varepsilon(y-2\pi k)
=\left\{\begin{array}{rl}
+1 &2n<\frac{y}{2\pi}<(2n+1)\\
0  &\frac{y}{2\pi}=n\\
-1 &2n+1<\frac{y}{2\pi} <(2n+2)
\end{array}\right.
\end{equation}
is an odd function that makes the prefactor periodic along the whole
real axis. The function $\widetilde{\eta}(y)$ is given by
\begin{equation}
\label{eta}
\widetilde{\eta}(y)=
\lim_{\kappa\rightarrow\infty}
\sum_{k=-\kappa}^\kappa
\arctan\
t_\rho\left(\Lambda\varepsilon(y-2\pi k)\right).
\end{equation}
Notice that the function (\ref{eta}) has the same shape as
$$\eta(y)\arctan t_\rho(\Lambda)$$
where $\eta(y)$ is the staircase function 
\begin{equation}
\label{beta2}
\eta(y)=
\lim_{\kappa\rightarrow\infty}
\sum_{k=-\kappa}^\kappa
\varepsilon(y-2\pi k)=\left\{
\begin{array}{cl}
2n & \frac{y}{2\pi}=n\\
2n+1 & n<\frac{y}{2\pi}<n+1
\end{array}
\right.
\end{equation}
that steps by two units every $2\pi$. However obviously both functions
differ by their derivatives on the brane (in the special case $\rho=1$
the two functions are indeed identical,
$\widetilde\eta(y)=\Lambda\eta(y)$).

It is a simple matter to prove that the functions (\ref{pereven}) are
periodic in the real axis $\mathbb{R}$ provided that condition
(\ref{periodiceven}) holds.

\section{\sc CP-odd brane mass terms}
The Lagrangian in this case is given by (\ref{bulk2}) and
(\ref{CPodd}), i.e.
\begin{equation}
\label{Lagodd}
\mathcal{L}_{\rm CP-odd}=i\,\bar\psi_i\gamma^\mu\partial_\mu \psi^i+
\bar\psi_2\partial_5\psi^1-\bar\psi_1\partial_5\psi^2-
\omega(\bar\psi_1\psi^1+\bar\psi_2\psi^2)-
2\Lambda\delta(y)\left[\bar\psi_1\gamma^5\psi^1-\rho\,\bar\psi_2
\gamma^5\psi^2\right]
\end{equation}
We can then write the EOM in the rest frame $p^\mu=(M_n,\vec{0})$
as
\begin{align}
\label{EOModd1}
\left(-M_n\gamma^0+\omega\right)\psi^1_n+\partial_y
\psi^2_n+2\phantom{\rho}\Lambda\delta(y)\gamma^5\psi^1_n=&0\nonumber\\
\left(-M_n\gamma^0+\omega\right)\psi^2_n-\partial_y
\psi^1_n-2\rho\,\Lambda\delta(y)\gamma^5\psi^2_n=&0\ .
\end{align}
where $\psi^i_n$ are the mass eigenstates corresponding to the mass
eigenvalues $M_n$. If we define now the Majorana spinors as in
(\ref{Majorana2}) we can write the EOM (\ref{EOModd1}) as
\begin{align}
\label{EOModd2}
-M_n\bar\phi^1_n+\omega\phi^1_n+\partial_y
\phi^2_n+2i\phantom{\rho}\Lambda\delta(y)\phi^1_n=&0\nonumber\\
-M_n\bar\phi^2_n+\omega\phi^2_n-\partial_y
\phi^1_n-2i\rho\,\Lambda\delta(y)\phi^2_n=&0
\end{align}

Unlike Eqs.~(\ref{EOM2}), Eqs.~(\ref{EOModd2}) do not admit real
solutions. In fact by decomposing
\begin{equation}
\label{decomp}
\phi^i_n=\varphi^i_n+i\,\chi^i_n
\end{equation}
Eqs.~(\ref{EOModd2}) lead to 
\begin{align}
\label{EOModd}
-(M_n-\omega)\varphi^1_n+\partial_y\, \varphi^2_n+
2\phantom{\rho}\Lambda \delta(y)\chi^1_n=&0 \nonumber\\
\phantom{-}(M_n+\omega)\chi^1_n+\partial_y\,\chi^2_n-
2\phantom{\rho}\Lambda \delta(y)\varphi^1_n=&0 \nonumber\\
-(M_n-\omega)\varphi^2_n-\partial_y\,\varphi^1_n-
2\rho\,\Lambda \delta(y)\chi^2_n=&0 \nonumber\\
\phantom{-}(M_n+\omega)\chi^2_n-\partial_y\,\chi^1_n+
2\rho\,\Lambda \delta(y)\varphi^2_n=&0 
\end{align}

Eqs.~(\ref{EOModd}) can be solved analytically.  We choose the most
general initial conditions consistent with parity (and up to a global
normalization factor):
\begin{align}
\label{initialodd}
\phi^1_n(0)=&\zeta, \nonumber\\
\chi^1_n(0)=&1,   \nonumber\\
\phi^2_n(0)=&0,   \nonumber\\
\chi^2_n(0)=&0. 
\end{align}
In the interval $-2\pi<y<2\pi$ an analytic solution is then given by
\begin{equation}
\label{soloddphi}
\left(
\begin{array}{c}
\phi^1_n(y) \vspace{.25cm}\\
\phi^2_n(y)
\end{array}
\right)=\sqrt{\frac{\zeta^2+t_\rho^2\left(\Lambda\varepsilon(y)\right)}
{1+\rho\ t_\rho^2\left(\Lambda\varepsilon(y)\right)}}
\left(
\begin{array}{c}
\cos\left[(M_n-\omega)y-
\arctan\left\{ t_\rho\left(\Lambda\varepsilon(y)\right)/\zeta \right\}\right]  \vspace{.25cm}\\ 
\sin\left[(M_n-\omega)y-
\arctan\left\{ t_\rho\left(\Lambda\varepsilon(y)\right)/\zeta \right\}\right]
\end{array}
\right)
\end{equation}
\begin{equation}
\label{soloddchi}
\left(
\begin{array}{c}
\chi^1_n(y) \vspace{.25cm}\\
\chi^2_n(y)
\end{array}
\right)=\sqrt{\frac{1+\zeta^2\, t_\rho^2\left(\Lambda\varepsilon(y)\right)}
{1+\rho\ t_\rho^2\left(\Lambda\varepsilon(y)\right)}}
\left(
\begin{array}{c}
\cos\left[(M_n+\omega)y-
\arctan\left\{ \zeta\,\,t_\rho\left(\Lambda\varepsilon(y)\right)\right\}\right] \vspace{.25cm} \\
-\sin\left[(M_n+\omega)y
-\arctan\left\{ \zeta\,\,t_\rho\left(\Lambda\varepsilon(y)\right)\right\}\right]
\end{array}
\right)
\end{equation}
where both $\zeta$ and $M_n$ are to be determined by imposing periodicity
of the solution.

Periodicity condition of (\ref{soloddphi}) and (\ref{soloddchi}) is
implemented by
\begin{equation}
\label{periododd}
M_n=n+\frac{1}{\pi}\,\arctan\sqrt{\frac{\tan^2(\omega\pi)+t_\rho^2(\Lambda)}
{1+\tan^2(\omega\pi)\,t_\rho^2(\Lambda)}}
\end{equation}
and the initial condition is fixed by
\begin{equation}
\label{initodd}
\zeta=\frac{\tan(M_n+\omega)\pi}{t_\rho(\Lambda)}=
\frac{t_\rho(\Lambda)}{\tan(M_n-\omega)\pi}
\end{equation}
where the last equality is implied by (\ref{periododd}) and $\zeta$
turns out to be independent of $n$. Notice that, unlike the case with
$CP$-even brane masses, when $CP$-odd brane masses are present
supersymmetry restoration is not possible for non-trivial values of
the parameters $\omega$ and $\Lambda$.

The solutions (\ref{soloddphi}) and (\ref{soloddchi}) near the brane,
$|y|<\epsilon$ where $\epsilon\to 0$, are given by,
\begin{equation}
\label{oddphibrane}
\left(
\begin{array}{c}
\phi^1_n(y) \vspace{.25cm}\\
\phi^2_n(y)
\end{array}
\right)=\frac{1}{\sqrt{
1+\rho\ t_\rho^2\left(\Lambda\varepsilon(y)\right)}}
\left(
\begin{array}{c}
\zeta \vspace{.25cm} \\
-t_\rho\left(\Lambda\varepsilon(y)\right)
\end{array}
\right)
\end{equation}
\begin{equation}
\label{oddchibrane}
\left(
\begin{array}{c}
\chi^1_n(y) \vspace{.25cm}\\
\chi^2_n(y)
\end{array}
\right)=\frac{1}{\sqrt{
1+\rho\ t_\rho^2\left(\Lambda\varepsilon(y)\right)}}
\left(
\begin{array}{c}
1  \vspace{.25cm}
\\
\zeta\,t_\rho\left(\Lambda\varepsilon(y)\right)
\end{array}
\right)
\end{equation}

From (\ref{oddphibrane}) and (\ref{oddchibrane}) we see that
$\phi^2_n$ and $\chi^2_n$ are odd functions that make jumps at the
brane $\Delta\phi^2_n$, $\Delta\chi^2_n$ given by
\begin{equation}
\label{jumpphi}
\left|\Delta\phi^2_n\right|=\frac{2 t_\rho(\Lambda)}{\sqrt{
1+\rho\ t_\rho^2\left(\Lambda\right)}}
\end{equation}
\begin{equation}
\label{jumpchi}
\left|\Delta\chi^2_n\right|=\frac{2\zeta\, t_\rho(\Lambda)}{\sqrt{
1+\rho\ t_\rho^2\left(\Lambda\right)}}
\end{equation}
while $\phi^1_n$ and $\chi^1_n$ have discontinuities on the brane
\begin{equation}
\label{disphi}
\lim_{y\to 0}\phi^1_n(y)=\frac{\zeta}{\sqrt{
1+\rho\ t_\rho^2(\Lambda)}}
\end{equation}
\begin{equation}
\label{discchi}
\lim_{y\to 0}\chi^1_n(y)=\frac{1}{\sqrt{
1+\rho\ t_\rho^2(\Lambda)}}
\end{equation}
Only for the particular case $\rho=0$ are the functions $\phi^1_n(y)$
and $\chi^1_n(y)$ continuous. However one can check that near the
brane
\begin{equation}
\label{contodd}
\lim_{y\to
0}\left[\phi^1_n(y)\,\chi^1_n(y)-\rho\,\phi^2_n(y)\,\chi^2_n(y)\right]=
\phi^1_n(0)\,\chi^1_n(0)-\rho\,\phi^2_n(0)\,\chi^2_n(0)
\end{equation}
and so the term coupled to the brane in the Lagrangian (\ref{Lagodd})
is continuous.

If we replace $\delta(y)\to \sum_k \delta(y-2\pi k)$ in (\ref{Lagodd})
the solution (\ref{soloddphi}) and (\ref{soloddchi}) can be
generalized to one periodic for $y\in\mathbb{R}$ by replacing
$\varepsilon(y)\to\beta(y)$ in the prefactors of (\ref{soloddphi}) and
(\ref{soloddchi}) and by replacing
\begin{align}
\label{replacing}
\arctan\, \zeta\, t_\rho[\Lambda\varepsilon(y)]\to&
\lim_{\kappa\rightarrow\infty}
\sum_{k=-\kappa}^\kappa
\arctan\, \zeta\, 
t_\rho[\Lambda\varepsilon(y-2\pi k)]\nonumber\\
\arctan\, \frac{1}{\zeta}\, t_\rho[\Lambda\varepsilon(y)]\to&
\lim_{\kappa\rightarrow\infty}
\sum_{k=-\kappa}^\kappa
\arctan\,  \frac{1}{\zeta}\,
t_\rho[\Lambda\varepsilon(y-2\pi k)]
\end{align}
in the corresponding arguments of the trigonometric functions.
Using (\ref{periododd}) and (\ref{initodd}) one can check that the new
functions are periodic along the real axis.

As in the case of $CP$-even mass terms we have checked our solution
numerically with a regularized $\delta$-function. We find perfect agreement
with our analytical solution, Eqs.~(\ref{soloddphi})--(\ref{initodd}).

\section{\sc Conclusions}

In this paper we have studied the compactification of a 5D theory on
$S^1/\mathbb{Z}_2$ with SS supersymmetry breaking and the most general
brane mass term. We have focussed on the case of bulk symplectic
Majorana fermions and so our results apply to the case of both
gauginos and gravitinos. In all cases fermions obey first order
differential equations and brane terms correspond to discontinuities
on the fields.

We have studied in full detail both the CP-even and CP-odd brane mass
cases. It has been shown that in the former case there can be a
cancellation between the contribution of the standard SS and the brane
terms and so supersymmetry can be restored whereas in the latter case
this is not possible. Although in most cases there is a discontinuous
behaviour for the even fermions near the brane, we have checked that
the combination that couples to the brane is continuous throughout the
orbifold fixed points. The generalization of this mechanism to a more
general framework with brane fields, or to warped scenarios is left
for future investigation.

\section*{\sc Acknowledgments}
We would like to thank J.~Bagger and M.~Redi for discussions. One of
us (AD) wants to thank the Aspen Center for Physics where part of this
work was done. GG thanks the DAAD for financial support.


\begin{thebibliography}{99}
%
%
\bibitem{ignas}
I.~Antoniadis, \PLB{246}{1990}{377};
%%CITATION = PHLTA,B246,377;%%
I.~Antoniadis, C.~Mu\~noz and M.~Quir\'os, \NPB{397}{1993}{515};
%%CITATION = HEP-PH 9211309;%%
I.~Antoniadis and K.~Benakli, \PLB{326}{1994}{69};
%%CITATION = HEP-TH 9310151;%%
I.~Antoniadis, K.~Benakli and M.~Quir\'os, \PLB{331}{1994}{313};
%%CITATION = HEP-PH 9403290;%%
I.~Antoniadis and M.~Quir\'os, \PLB{392}{1997}{61}.
%%CITATION = HEP-TH 9609209;%%
%
%
\bibitem{SS}
J.~Scherk and J.H.~Schwarz, \PLB{82}{1979}{60}; \NPB{153}{1979}{61};
%%CITATION = PHLTA,B82,60;%%
%%CITATION = NUPHA,B153,61;%%
P.~Fayet, \PLB{159}{1985}{121}; \NPB{263}{1986}{649}.
%%CITATION = PHLTA,B159,121;%%
%%CITATION = NUPHA,B263,649;%%
%
%
\bibitem{matter}
A.~Pomarol and M.~Quir\'os, \PLB{438}{1998}{255};
%%CITATION = HEP-PH 9806263;%%
I.~Antoniadis, S.~Dimopoulos, A.~Pomarol and M.~Quir\'os,
\NPB{544}{1999}{503};
%%CITATION = HEP-PH 9810410;%%
A.~Delgado, A.~Pomarol and M.~Quir\'os, \PRD{60}{1999}{095008}.
%%CITATION = HEP-PH 9812489;%%
%

\bibitem{UV}
R.~Barbieri, L.J.~Hall and Y.~Nomura, \PRD{63}{2001}{105007};
%%CITATION = HEP-PH 0011311;%%
N.~Arkani-Hamed,  L.J.~Hall, Y.~Nomura, D.~Smith and N.~Weiner,
\NPB{605}{2001}{81};
%%CITATION = HEP-PH 0102090;%%
A.~Delgado, and M.~Quir\'os, \NPB{607}{2001}{99}.
%%CITATION = HEP-PH 0103058;%%
%
\bibitem{SSH1}
G.~von Gersdorff and M.~Quiros,
\PRD{65}{2002}{064016}.
%%CITATION = HEP-TH 0110132;%%
%
\bibitem{SSH2}
G.~von Gersdorff, M.~Quiros and A.~Riotto,
\NPB{634}{2002}{90}.
%%CITATION = HEP-TH 0204041;%%
%
\bibitem{radionF}
Z.~Chacko and M.A.~Luty, \JHEP{0105}{2001}{067};
%%CITATION = HEP-PH 0008103;%%
T.~Kobayashi and K.~Yoshioka,
\PRL{85}{2000}{5527};
%%CITATION = HEP-PH 0008069;%%
D.~Mart\'{\i} and A.~Pomarol, 
{\it Phys.\ Rev.} {\bf D64} (2001) 105025;
%%CITATION = HEP-TH 0106256;%%
D.E.~Kaplan and N.~Weiner, 
\texttt{hep-ph/0108001}.
%%CITATION = HEP-PH 0108001;%%
%
\bibitem{one}
A.~Delgado, G.v.~Gersdorff, P.~John and M.~Quir\'os, \PLB{517}{2001}{445};
%%CITATION = HEP-PH 0104112;%%
R.~Contino, L.~Pilo, R.~Ratazzi and A.~Strumia, \JHEP{0106}{2001}{005};
%%CITATION = HEP-PH 0103104;%%
R.~Contino and L.~Pilo, \PLB{523}{2001}{347};
%%CITATION = HEP-PH 0104130;%%
Y.~Nomura, \texttt{hep-ph/0105113};
%%CITATION = HEP-PH 0105113;%%
N.~Weiner, \texttt{hep-ph/0106021}; 
%%CITATION = HEP-PH 0106021;%%
A.~Delgado, G.v.~Gersdorff and M.~Quir\'os, \NPB{613}{2001}{49};
%%CITATION = HEP-PH 0107233;%%
V.D.~Clemente, S.F.~King and D.A.~Rayner, \NPB{617}{2001}{71};
%%CITATION = HEP-PH 0107290;%%
R.~Barbieri, L.J.~Hall and Y.~Nomura, \NPB{624}{2002}{63};
%%CITATION = HEP-TH 0107004;%%
V.D.~Clemente and Y.A.~Kubyshin, \NPB{636}{2002}{115};
%%CITATION = HEP-TH 0108117;%%
H.~D.~Kim,
%``Softness of Scherk-Schwarz supersymmetry breaking,''
Phys.\ Rev.\ D {\bf 65} (2002) 105021.
%%CITATION = HEP-TH 0109101;%%
%
\bibitem{orbifolds}
L.~J.~Dixon, J.~A.~Harvey, C.~Vafa and E.~Witten,
\NPB{261}{1985}{678};
%%CITATION = NUPHA,B261,678;%%
%
L.~J.~Dixon, J.~A.~Harvey, C.~Vafa and E.~Witten,
\NPB{274}{1986}{285}.
%%CITATION = NUPHA,B274,285;%%
%
\bibitem{AQ}
I.~Antoniadis and M.~Quiros,
\NPB{505}{1997}{109};
%%CITATION = HEP-TH 9705037;%%
%
I.~Antoniadis and M.~Quiros,
\PLB{416}{1998}{327};
%%CITATION = HEP-TH 9707208;%%
%
I.~Antoniadis and M.~Quiros,
Nucl.\ Phys.\ Proc.\ Suppl.\  {\bf 62} (1998) 312.
%%CITATION = HEP-TH 9709023;%%
%
\bibitem{fabio1}
J.~A.~Bagger, F.~Feruglio and F.~Zwirner,
\PRL{88}{2002}{101601}.
%%CITATION = HEP-TH 0107128;%%
%
\bibitem{fabio2}
J.~Bagger, F.~Feruglio and F.~Zwirner,
\JHEP{0202}{2002}{010};
%%CITATION = HEP-TH 0108010;%%
T.~Gherghetta and A.~Riotto,
%``Gravity-mediated supersymmetry breaking in the brane-world,''
{\it Nucl.\ Phys.} {\bf B623} (2002) 97.
%%CITATION = HEP-TH/0110022;%%
%
\bibitem{nilles}
K.~A.~Meissner, H.~P.~Nilles and M.~Olechowski,
\texttt{hep-th/0205166}.
%%CITATION = HEP-TH 0205166;%%
%
\bibitem{feruglio}
C.~Biggio and F.~Feruglio,
%``Symmetry breaking for bosonic systems on orbifolds,''
{\it Annals Phys.}  {\bf 301} (2002) 65;
%%CITATION = HEP-TH 0207014;%%
C.~Biggio, F.~Feruglio, A.~Wulzer and F.~Zwirner,
\texttt{hep-th/0209046}.
%%CITATION = HEP-TH 0209046;%%
%
%
\bibitem{zucker}
M.~Zucker, \NPB{570}{2000}{267};
%%CITATION = HEP-TH 9907082;%%
%
\JHEP{08}{2000}{016};
%%CITATION = HEP-TH 9909144;%%
%
\PRD{64}{2001}{024024}.
%%CITATION = HEP-TH 0009083;%%
%

\bibitem{kugo}
T.~Kugo and K.~Ohashi,
%``Supergravity tensor calculus in 5D from 6D,''
{\it Prog.\ Theor.\ Phys.}  {\bf 104} (2000) 835;
%%CITATION = HEP-PH 0006231;%%
T.~Kugo and K.~Ohashi,
%``Off-shell d = 5 supergravity coupled to matter-Yang-Mills system,''
{\it Prog.\ Theor.\ Phys.}  {\bf 105} (2001) 323;
%%CITATION = HEP-PH 0010288;%%
T.~Fujita, T.~Kugo and K.~Ohashi,
%``Off-shell formulation of supergravity on orbifold,''
{\it Prog.\ Theor.\ Phys.} \textbf{106} (2001) 671;
%%CITATION = HEP-TH 0106051;%%
T.~Kugo and K.~Ohashi,
%``Superconformal Tensor Calculus on Orbifold in 5D,''
{\it Prog.\ Theor.\ Phys.} {\bf 108}, 203 (2002).
%%CITATION = HEP-TH 0203276;%%


\end{thebibliography}
\end{document}